# HELIOSHEATH SYNCHROTRON RADIATION AS A POSSIBLE SOURCE FOR THE ARCADE 2 CMB DISTORTIONS

H.N. Sharpe

Bognor, Ontario, Canada
sh3149@brucetelecom.com

## ABSTRACT

This brief note speculates that the recently reported residual CMB signal [Seiffert et al 2009] may originate within the Sun's heliosheath. A temperature spectrum function is derived that has the same power law form as the fitted function in Seiffert et al. In particular a spectral index of +2 is implied. An optically thin radiating shell of thickness ~ 1AU could match the required 1K deg power law amplitude. A possible mechanism for the heliosheath magnetic fields is discussed based on Alfven's heliospheric current model with embedded double layers as the energy source for the relativistic electrons.

## 1 INTRODUCTION

Recently Seiffert et al [2009] have presented data from the ARCADE 2 instrument that suggests a residual signal beyond the CMB background after correcting for potential contributions from various extra-galactic sources. The spectrum function they fit to the ARCADE (and other) data has the following form:

$$T(\nu) = T_0 + A\,(\nu/1\text{GHz})^{\beta} + \Delta T(\nu) \tag{1}$$

where $T_0$ is the CMB baseline temperature, A is the power law amplitude at 1 GHz, $\beta$ is the power law index, $\nu$ is the frequency and $\Delta T(\nu)$ is a CMB spectral distortion. They find that the unexplained residual emission is consistent with a power law amplitude $A = 1.06 \pm 0.11$ K at 1 GHz and a spectral index $\beta = -2.56 \pm 0.04$.

It is the purpose of this brief note to speculate that the residual emission may in fact originate from an optically thin layer within the heliosheath which surrounds the solar system. The energetic electrons follow a power law spectrum in the presence of random magnetic fields. In this idealized model, the synchrotron radiation is homogeneous and isotropic. Polarization is not to be expected. The first part of this note develops the relevant temperature spectrum formula. The second part discusses values for the parameters and possible mechanisms.

## 2 EQUIVALENT TEMPERATURE SPECTRUM

The observed brightness of a background source, $I_{source}$, through a foreground shell with intrinsic brightness, $I_{shell}$, and optical depth $\tau$ is [Rybicki and Lightman 1979]:

$$I_{obs} = I_{source}\,e^{-\tau} + I_{shell}\,(1 - e^{-\tau}) \tag{2}$$

We assume spherical symmetry. For an optically thin shell ($\tau \ll 1$):

$$I_{obs} = I_{source} + I_{shell}\,\tau \tag{3}$$

For $\tau$ constant across the shell thickness L,

$$I_{obs} = I_{source} + L\, j_{shell} \tag{4}$$

where $j_{shell}$ is the specific emissivity for the radiating shell. Assume a power law energy distribution for the radiating electrons with spectral index p:

$$N(E)dE = K\, E^{-p}\, dE \tag{5}$$

Then the observed brightness is:

$$I_{obs} = I_{source} + 1.35 \times 10^{-22}\, K\, B^{(p+1)/2}\, a(p) L\, (6.26 \times 10^{18} / \nu)^{(p-1)/2} \tag{6}$$

where B is the magnetic field and a(p) is a function used to simplify the power law index dependency [Ginzburg and Syrovatskii 1965]. The units of I are erg cm$^{-2}$ sec$^{-1}$ ster$^{-1}$ Hz$^{-1}$. K has units ergs$^{p-1}$cm$^{-3}$.

We convert to the equivalent temperature using the Rayleigh-Jeans relation: $T_{eq}=(c^2/2k\nu^2)I$, where k is the Boltzmann constant:

$$T_{obs} = T_{source} + [\, 4.4 \times 10^{14}\, K\, L\, a(p)\, B^{(p+1)/2}\, (6.26 \times 10^{18})^{(p-1)/2}\,]\, \nu^{-(p+3)/2} \tag{7}$$

At this point we identify $T_{source}$ with $T_{zero} + \Delta T(\nu)$ in (1) and focus on the temperature spectrum contribution from the optically thin intervening shell.

The power law index $\beta$ from the ARCADE data fit implies a spectral index for the radiating electrons of p = +2 (from (p+3)/2 = 2.5). A value of +2 is typical for relativistic electrons assuming an isotropic pitch angle distribution. The corresponding value for a(p) is 0.103 [Burke and Graham-Smith 2002].

Using these values in (7) for the shell effective temperature spectrum we obtain an observed effective temperature for the shell ($\nu$ in GHz):

$$T_{shell} = [\, 3.48\, K\, L\, B^{3/2}\,]\, \nu_{GHz}^{-2.5} \tag{8}$$

Integrating (5) and using p = 2, K is seen to be an electron energy density for this case:

$$K = n_0 E_1 \tag{9}$$

where $n_0$ is the electron number density in the shell and $E_1$ is the lower bound on the energy interval. We identify $E_1$ with the critical electron energy $E_{cr}$, associated with the critical frequency. Finally we have the desired functional form that corresponds to the residual emission in (1):

$$T_{shell} = A\, \nu_{GHz}^{-2.5} \tag{10}$$

where $A \equiv 3.48\, n_0\, E_{cr}\, L\, B^{3/2}$. We wish to investigate the conditions under which the value of A equals the fitted power law amplitude value of 1 Kdeg in (1). The four physical factors which contribute to A are the electron number density, the electron energy that corresponds to the critical frequency, the magnetic field and the shell thickness. As a first pass we compute the required value of L given representative values of the heliosheath parameters. Using $n_0 = 0.1\, cm^{-3}$, $B = 1\, \mu G$ [Peratt 1992; Burlaga et al. 2006] and $E_{cr} = 1$ Gev which corresponds to a critical frequency of 16MHz, we compute L ~ 1 AU for A = 1Kdeg. The heliosheath is expected to be much larger than this value [Czechowski et al 2006]. The essential point is that the required thickness of the optically thin shell is not inconsistent with the power law amplitude of 1K deg.

## 3 MECHANISMS

Next we consider the nature of the heliosheath magnetic fields and the energetic electrons. The heliosheath is a highly complex and not well-understood region of the heliosphere. While the Voyager spacecraft are improving this understanding, the discussion which follows is necessarily speculative. Alfven [1981] developed a heliospheric current model along the lines of the observed terrestrial magnetospheric current system. Briefly, the Sun acts as a unipolar inductor that generates an emf which drives two polar current systems. These currents spread out, possibly to the heliosheath, and return to the Sun near the equatorial plane thereby completing the "circuit". Observational data supporting this model are presented in Peratt [1992]. It is not known if the polar currents spread out to form current sheets or if they concentrate into filaments. It is likely that they break up into cylindrical filaments and pinch down in a Bennett-type pinch effect. But whereas the Bennett pinch represents a balance between the plasma pressure and a compressing electromagnetic force, a low density plasma like that in the heliosheath cannot generate an internal plasma pressure. The result is a force-free magnetic field in a reference frame which moves with the current because the electric and magnetic fields are aligned in this frame. To an outside observer the magnetic field would appear as a twisted magnetic "rope" [Alfven and Falthammar 1963]. Force-free fields represent the lowest state of magnetic energy that a closed system may attain [Peratt 1992]. Hence an initial current sheet should break into Birkeland (magnetic field aligned) vortex tubes having a characteristic distribution function. Recent Voyager data suggest the presence of magnetic rope structures in the heliosheath [Burlaga et al 2006].

We may obtain an estimate for the filament radius in a force-free situation following Carlqvist [1988]:

$$a = (\mu_0 I_p) / (2\pi M B_z) \qquad (\mu_0: \text{free space permeability}) \quad (11)$$

where $I_p$ is the solar supply polar current, $B_z$ is the axial magnetic field in the heliosheath (Tesla), and M is the number of filaments. For a supply solar current of 1.5x10**9 A [Alfven 1981] we have: $a = 3 \times 10^{**}8/(MB_z)$ (a in cm and B in gauss). For $B_z = 1$ $\mu$G and a = 1 AU we would have about 100 such vortex tubes spreading out from the Sun's polar axis into each hemisphere of the heliosheath. It is likely that such large filaments would continue to break up into smaller tubes thereby generating a spectrum of size scales and enhanced axial magnetic fields ( since $B_z$ scales as $1/a^2$).

Relativistic electrons in the heliosheath could be produced in a system of double layers distributed along the field-aligned current systems. The double layer is an electrostatic structure a few Debye lengths wide which may appear in a current carrying plasma. It can sustain a high net potential difference which could accelerate electrons traversing it. It may then provide a mechanism for transforming stored magnetic energy into the directed kinetic energy of the accelerated particles [Peratt 1992]. Alfven [1986] discusses the production of relativistic electrons in double layers associated with a heliospheric current system.

## 4 DISCUSSION

This note has presented a speculation for the residual emission reported by Seiffert et al. By postulating an optically thin low density magnetized plasma shell, possibly associated with the Sun's heliosheath, we derived an equation for the observed effective shell temperature spectrum consistent with that reported for the residual emission. We have also suggested a possible mechanism for this radiation in terms of a heliospheric current system. However the validation of this mechanism is dificult. In particular it must be shown that a power spectral index of +2 can actually be derived for the radiating electrons in a distribution of magnetized vortex tubes. It must also be shown that the resultant radiation is isotropic, though it may contain higher moments due to distortion of the heliosheath by the local ISM. Finally improved observational and theoretical constraints are required for the parameters that comprise the power law amplitude in (10). Overlying all these problems is the need to better understand heliosheath physics and whether the heliosheath can in fact support a global current system. A correlation between the residual emission and the solar cycle could suggest this support.